\documentclass[aps,prmaterials,reprint,amsmath,amssymb]{revtex4-2}
\usepackage[utf8]{inputenc}
\usepackage{graphicx}
\usepackage{booktabs}
\usepackage{xurl}
\usepackage{hyperref}
\hypersetup{colorlinks=true,linkcolor=blue,citecolor=blue,urlcolor=blue}
\newcommand*\chem[1]{\ensuremath{\mathrm{#1}}}
\begin{document}
\title{Thermoelectric properties and modulation doping of ultrawide-bandgap nitride monolayers}
\author{Ariel R. Barr}
\email{arbarr@mit.edu}
\affiliation{Department of Materials Science \& Engineering, Massachusetts Institute of Technology, Cambridge, Massachusetts 02139, USA}
\begin{abstract}
Two-dimensional III-nitride alloys provide a composition-tunable platform
for exploring thermoelectric transport and intrinsic electric-field
tolerance. We combine first-principles electronic-structure calculations
with semiclassical transport for ordered monolayer \chem{B_xAl_{1-x}N}
and \chem{B_xGa_{1-x}N}. State-dependent momentum relaxation times
computed with AMSET include acoustic deformation-potential and piezoelectric
scattering. Using a common, temperature-independent lattice thermal
conductivity of $550~\mathrm{W\,m^{-1}\,K^{-1}}$, we predict maximum
$ZT>2.5$ for selected p- and n-type monolayers and $ZT>2$ at
$300~\mathrm K$ for n-type \chem{B_{0.25}Ga_{0.75}N}.
Composition-dependent band curvature, screening, and scattering help
interpret the different p- and n-type responses. Vacuum-referenced band
alignments identify candidate modulation-doping partners for both carrier
types; explicit hGaN/\chem{V_2O_5} calculations yield a hole surface
density of approximately $1.6\times10^{14}~\mathrm{cm^{-2}}$.
A separate phenomenological model using the band gap and maximum phonon
frequency predicts intrinsic breakdown fields of $20.06$ and
$20.74~\mathrm{MV/cm}$ for the Al- and Ga-containing $x=0.75$ alloys,
respectively. These results motivate further investigation for
thermoelectric and high-field power-electronic applications. The $ZT$
values apply to the specified scattering and lattice-conductivity models,
while the breakdown estimates require validation for monolayers and do not
resolve field direction.
\end{abstract}
\maketitle

\section{Introduction}

III-nitride semiconductors combine wide bandgaps with chemical, mechanical,
and thermal stability~\cite{ambacher1998growth,zhang2022multidimensional},
supporting applications in high electron mobility
transistors~\cite{huang2024threshold,chandrasekar2020demonstration},
light-emitting diodes~\cite{li2018algan},
lasers~\cite{nakamura1997high,yoshida2008demonstration}, and quantum
sensors~\cite{gottscholl2021spin,gilardoni2024single}.
The high critical fields of wide-bandgap materials such as GaN motivate
their use in power switching and RF electronics~\cite{shenai1989optimum,tan2015emerging}.
Ultrawide-bandgap (UWBG) materials extend this search to gaps wider than
that of GaN, including AlN, Al-rich AlGaN, $\beta$-\chem{Ga_2O_3}, and
diamond~\cite{tsao2018ultrawide,wong2021ultrawide}.
Their prospects depend jointly on field tolerance, carrier transport,
dopability, contacts, and heat removal. These challenges differ by material:
low thermal conductivity is a particular limitation of
$\beta$-\chem{Ga_2O_3}, whereas diamond has exceptionally high thermal
conductivity~\cite{tsao2018ultrawide}.
Monolayer III-nitride alloys offer an additional setting in which
composition and reduced dimensionality can modify the electronic
structure and intrinsic field tolerance.

Thermoelectric energy conversion provides a complementary motivation for
studying these alloys. Waste-heat recovery and localized cooling require
materials that balance electrical conduction and thermopower against heat
conduction~\cite{wang2012fluid,schierning2018bring,zebarjadi2012perspectives}.
\chem{Bi_2Te_3}-based alloys are established low-temperature
thermoelectrics~\cite{hao2016high,gayner2016recent,poudel2008high}, while
device-integrated thermoelectric cooling offers a route to managing local
heat generation~\cite{liu2024thermoelectric}.
Large bandgaps increase the energy required for thermal electron--hole
pair generation and can help suppress bipolar transport at elevated
temperatures. A wide gap alone, however, does not establish a large
thermoelectric figure of merit: the band dispersion, carrier scattering,
and lattice heat transport must also be considered. Low-temperature
transport is particularly relevant to wearable generators and
sensors~\cite{yang2023flexible,zhang2022micro,miao2024comfortable,jia2021wearable,ryu2019hybrid,zhu2022review}.

The tellurium-free composition of the III-nitrides also motivates
examining their materials cost and suitability for low-toxicity
applications. The historical laboratory-grade \chem{Bi_2Te_3} catalog
prices cited here range from \$6 to \$17 per
gram~\cite{Bi2Te3price1,Bi2Te3price2}, compared with constituent-material
estimates of approximately \$0.12/g for Ga-based nitrides and \$0.003/g
for Al-based nitrides~\cite{Bprice,Nprice,Gaprice}.
These estimates motivate a possible material-cost advantage; they exclude
monolayer synthesis, processing, and device fabrication.
Reports of biocompatibility and low cytotoxicity for GaN
surfaces~\cite{jewett2012ganbiocompatibility} provide further motivation,
while the biocompatibility of each alloy and device structure requires
separate evaluation.

Compared with bulk, nanowire, and quantum-dot III-nitrides, their
two-dimensional forms provide distinct bonding environments and
opportunities for interface control~\cite{arora2024ordered}.
In boron-based ordered monolayers, composition-induced puckering changes
the band-edge orbital character, symmetry, and directional dispersion.
Band curvature and dielectric screening therefore offer useful
descriptors for interpreting composition-dependent transport, while
vacuum-referenced band edges guide the selection of contacting materials.
Two-dimensional interfaces also permit modulation doping, in which an
adjacent layer supplies carriers to the conducting
layer~\cite{wang2021modulation,arora2025engineering}.
Spatial separation of the dopant and channel can reduce ionized-impurity
scattering, motivating its use for both electronic and thermoelectric
transport~\cite{mcclellan2021high,pei2014high,chen2022modulation,wang2022highly}.

Building on our studies of ordered nitride
monolayers~\cite{arora2024ordered} and interfacial charge-transfer
doping~\cite{arora2025engineering}, we investigate monolayer
\chem{B_xGa_{1-x}N} and \chem{B_xAl_{1-x}N} using first-principles
electronic structure and semiclassical transport theory.
Earlier calculations by Wines, Ersan, and Ataca examined transport
coefficients of related two-dimensional III-nitride alloys using a
constant relaxation-time approximation, but did not report
$ZT$~\cite{wines2020engineering}.
Here, state-dependent AMSET momentum relaxation
times~\cite{ganose2021scattering} are used to evaluate thermoelectric
response across temperatures and carrier concentrations, with particular
attention to performance near $300~\mathrm K$ under a stated common
lattice-conductivity reference. We interpret the results using the
composition-dependent masses, dielectric response, work functions, and
effective polar-optical-phonon frequencies in Table~\ref{tab:properties}.
Candidate p- and n-type modulation-doping partners are selected by band
alignment, with an explicit hGaN/\chem{V_2O_5} interface illustrating
hole transfer. We also estimate intrinsic breakdown fields with the
Kim--Pilania--Ramprasad band-gap/maximum-phonon-frequency
model~\cite{kim2016breakdown} and examine their relevance to high-field
power electronics. This maximum-frequency input is distinct from the
effective polar-optical frequency used to discuss carrier scattering.
The thermoelectric predictions, proposed carrier-control routes, and
intrinsic-field estimates are assessed within their respective modeling
assumptions.

\section{Results and discussion}
\label{sec:results}
\subsection{Crystal structure and electronic band structure}
\label{sec:materials}
We considered monolayer boron-based III-nitride ordered alloys, \chem{B_xGa_{1-x}N} and \chem{B_xAl_{1-x}N}, for boron fractions ranging from x = 0 to x = 1. Thermodynamic and dynamic stability of these materials has been demonstrated in a prior work of the author \cite{arora2024ordered}.
\begin{figure*}[!t]
    \centering
    \includegraphics[width=0.99\linewidth]{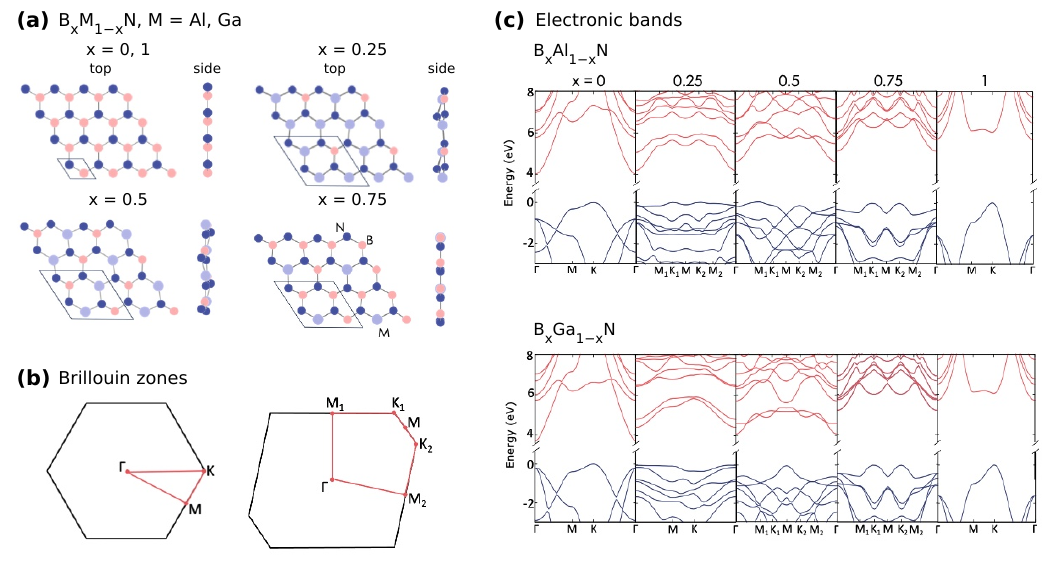}
    \caption{
    {\bf (a)} Top and side views of the 2D III-nitride crystal structures, arranged left to right and top to bottom for $x=0,1$; $0.25$; $0.5$; and $0.75$.
    {\bf (b)} The Brillouin zone of the hexagonal materials (left) and the reduced symmetry Brillouin zone of the puckered materials (right).
    {\bf (c)}  Electronic structure of  \chem{B_xAl_{1-x}N} (top), and \chem{B_xGa_{1-x}N} (bottom) computed using the hybrid DFT functional HSE06.
} \label{fig:band}
\end{figure*}

The evolution of the electronic structure with change in fractional B composition is shown in Fig.~\ref{fig:band}(c). When boron is introduced into pristine hexagonal GaN and AlN, the crystal structures buckle in the out-of-plane direction, with N atoms shifting from a trigonal-planar to a pyramidal bonding configuration, as shown in Fig.~\ref{fig:band}(a). This structure is induced by the large difference in atomic size between B and Al/Ga (atomic radii: 0.85 \AA \, and 1.25/1.30 \AA, respectively), which leads to a strained heterogeneous bonding environment that is relieved by adopting a lower energy puckered configuration. As a result, the lattice symmetry is reduced, permitting previously symmetry-forbidden mixing between $p_z$ and $p_x$, $p_y$ orbitals and modifying the valence-band character and bandgap. The resulting valence bands can exhibit strongly anisotropic dispersion and extrema at low-symmetry points of the Brillouin zone. Table~\ref{tab:properties} quantifies the band curvatures and electrostatic response; their connection to transport and charge transfer is discussed in Sec.~\ref{sec:electronic-response}.

\subsection{Thermoelectric performance}
\label{sec:thermoelectric}

\begin{figure*}[!tp]
    \centering
    \includegraphics[width=0.99\linewidth]{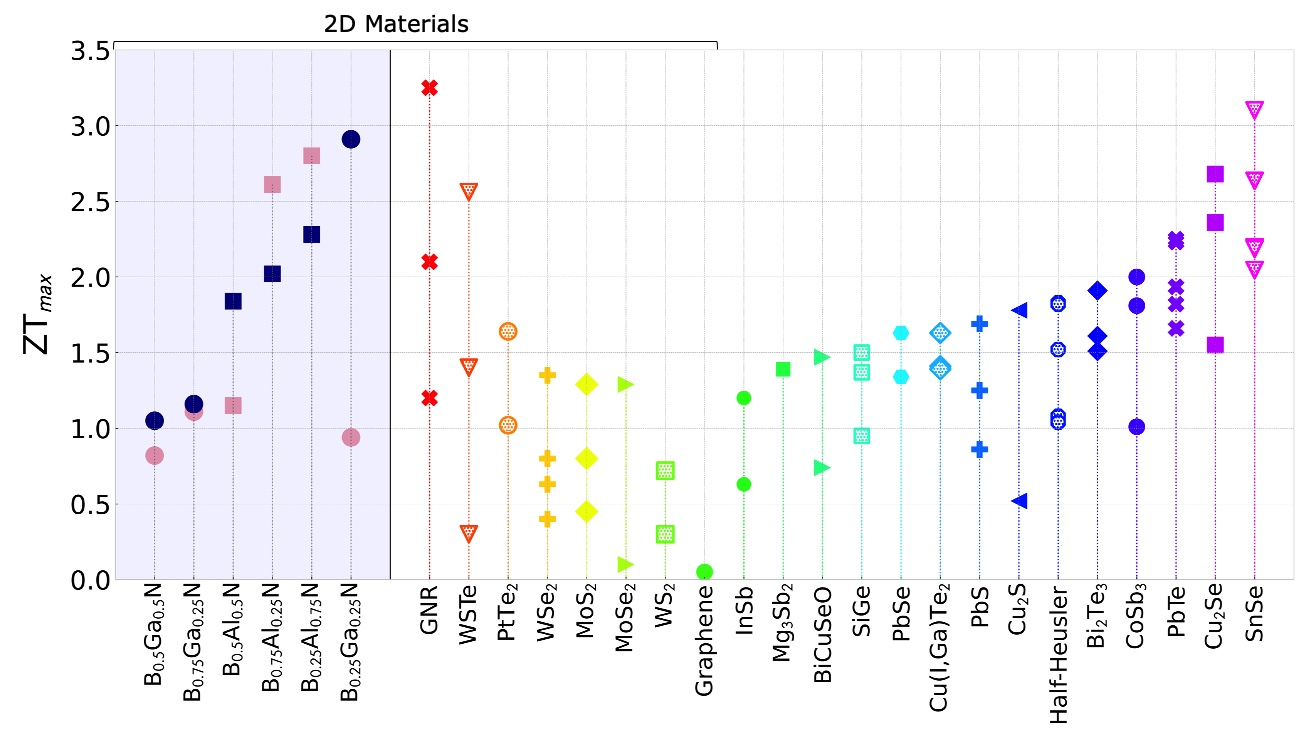}
    \caption{ Maximum thermoelectric figure of merit ZT$_{\rm max}$ of III-nitride monolayers compared with that of representative families of TE materials. The III-nitride estimates use $\kappa_{\mathrm l}=550~\mathrm{W\,m^{-1}\,K^{-1}}$ throughout; this assumption does not apply to the literature benchmarks. Other 2D materials include: graphene nano-ribbons (GNR) \cite{tran2017optimizing,sevinccli2013bottom,dollfus2015thermoelectric},  \chem{WS_{2}}
    \cite{li2020recent,bilc2021electronic,huang2013thermoelectric}, \chem{MoSe_{2}} \cite{kumar2015thermoelectric,li2020recent,huang2013thermoelectric}, \chem{MoS_{2}}
    \cite{liu2024polarization,wickramaratne2014electronic,huang2013thermoelectric,li2020recent,bilc2021electronic},  \chem{WSe_{2}} \cite{li2020recent,bilc2021electronic,kumar2015thermoelectric,huang2013thermoelectric}, \chem{PtTe_{2}}
    \cite{zhang2020remarkable,marfoua2019high},
     graphene \cite{d2018enhancement}, while
    bulk materials include: InSb \cite{cheng2017new,yamaguchi2005thermoelectric}, \chem{Mg_3Sb_{2}}  \cite{zhang2017discovery}, BiCuSeO \cite{liu2012copper,liu2013ultrahigh,zhong2014high,olvera2017partial}, SiGe \cite{wang2008enhanced,yu2012enhancement,bathula2012enhanced}, PbSe \cite{zhao2013high,wang2012weak}, Cu(I, Ga)Te$_2$ \cite{plirdpring2012chalcopyrite,luo2016progressive,zhang2014high}, PbS \cite{zhao2011high,wang2013high,ibanez2016high}, Cu$_2$S \cite{liu2012copper,liu2013ultrahigh,zhong2014high,olvera2017partial}, Half-Heusler \cite{kozinsky2021thermoelectrics,samsonidze2018accelerated,joshi2011enhancement,yan2013thermoelectric,fu2015realizing} , \chem{Bi_2Te_{3}}
    \cite{xie2009unique,kim2015dense,poudel2008high}, \chem{CoSb_{3}}
    \cite{sales1996filled,shi2011multiple,rogl2014n}, SnSe
    \cite{zhao2014ultralow,zhao2016ultrahigh,peng2016broad,duong2016achieving}.   N-type (p-type) doping of the 2D III-nitrides is indicated by navy (pink) markers respectively.
}\label{fig:ZT}
\end{figure*}

The thermoelectric performance of a material at a given temperature can be assessed by the dimensionless figure of merit
\begin{equation}
    \text{ZT} = \frac{\sigma \mathrm{S}^2 \mathrm{T}}{\kappa_{\mathrm{e}}+\kappa_{\mathrm{l}}}
\end{equation}
where $\sigma$ is the electrical conductivity, the Seebeck coefficient S, the temperature T, the electronic thermal conductivity $\kappa_{\mathrm{e}}$, and the lattice thermal conductivity $\kappa_{\mathrm{l}}$.

Carrier transport is evaluated using state-dependent AMSET MRTA lifetimes
for the stated acoustic deformation-potential and piezoelectric scattering
model~\cite{ganose2021scattering}. All nitride $ZT$ estimates use the common
reference $\kappa_{\mathrm l}^{\mathrm{ref}}=550~\mathrm{W\,m^{-1}\,K^{-1}}$,
held fixed across compositions and sampled temperatures. The models and normalization conventions are detailed in
Sec.~\ref{sec:methods}.

Figure~\ref{fig:ZT} presents the maximum thermoelectric figure of merit of the III-nitride monolayers and compares these predictions with representative thermoelectric material families. At all B-fractions a ZT$_{\rm max}$ $>$ 1 is exhibited, and a majority display an average ZT$_{\rm max}$ $>$ 1 on the temperature range 300-900K. Three monolayers exhibit exceptional high-temperature ZT$_{\rm max}$ $>$ 2.5, which compares favorably with both established TE materials such as \chem{Bi_2Te_{3}} and other emerging 2D materials such as transition metal dichalcogenides (TMDs). The composition-dependent balance between band dispersion, screening, and scattering underlying these results is examined using Table~\ref{tab:properties} in Sec.~\ref{sec:electronic-response}.

\subsection{Low-temperature thermoelectric performance}
\label{sec:low-temperature}

\begin{figure*}[!tp]
    \centering
    \includegraphics[width=0.99\linewidth]{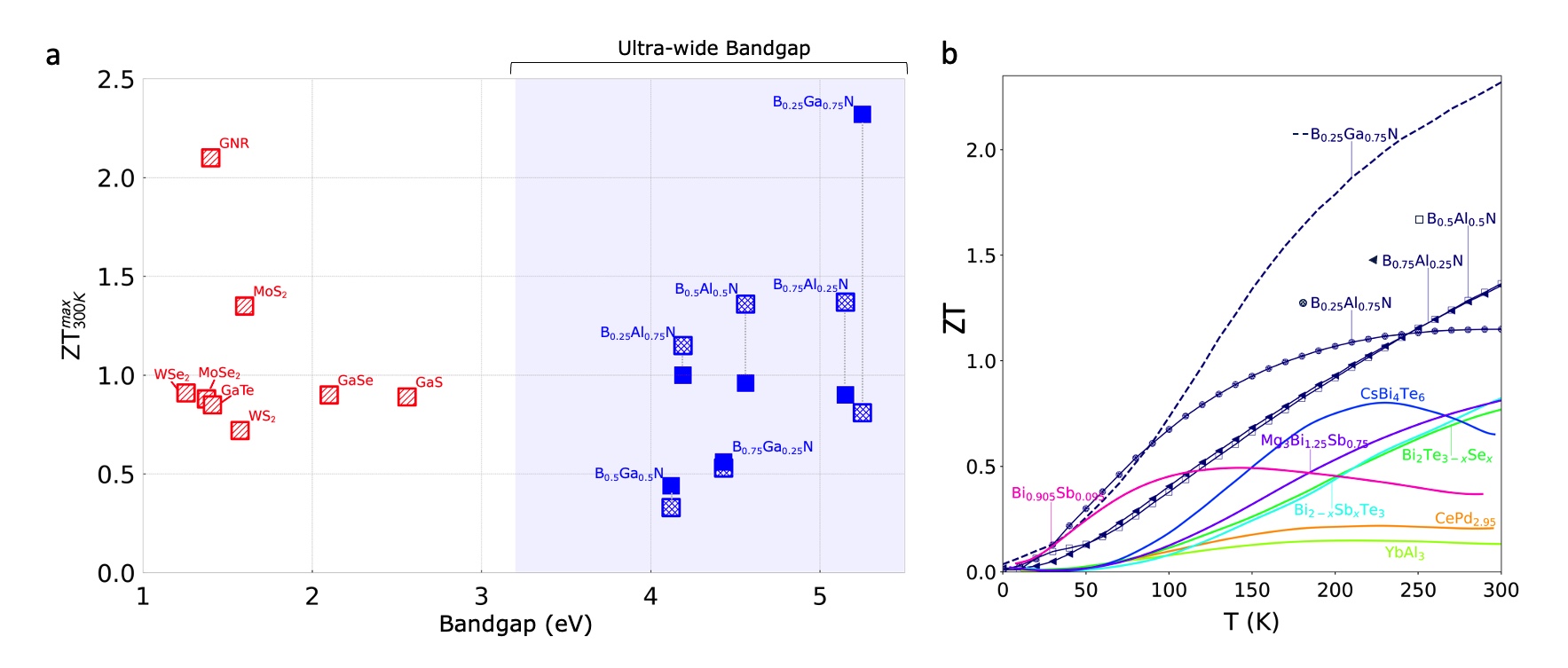}
    \caption{Maximum low-temperature ($300~\mathrm K$) thermoelectric figure of merit ZT$_{\rm 300K}$. The III-nitride estimates use $\kappa_{\mathrm l}=550~\mathrm{W\,m^{-1}\,K^{-1}}$ throughout; the literature benchmarks retain their original thermal-conductivity models or measurements. {\bf (a)} Comparison of monolayer 2D materials, including graphene \cite{d2018enhancement}, \chem{MoS_{2}}
    \cite{liu2024polarization,wickramaratne2014electronic,huang2013thermoelectric,li2020recent,bilc2021electronic}, \chem{MoSe_{2}} \cite{kumar2015thermoelectric,li2020recent,huang2013thermoelectric}, \chem{WS_{2}} \cite{li2020recent,bilc2021electronic,huang2013thermoelectric},  \chem{WSe_{2}} \cite{li2020recent,bilc2021electronic,kumar2015thermoelectric,huang2013thermoelectric},
    GaS \cite{bahuguna2018hybrid},
    GaSe \cite{bahuguna2018hybrid},
    GaTe \cite{bahuguna2018hybrid},
    and graphene nano-ribbons (GNR) \cite{tran2017optimizing,sevinccli2013bottom,dollfus2015thermoelectric}.
     ZT$_{\rm 300K}$ corresponding to n-type (p-type) 2D III-nitrides is indicated by solid (hatched) blue markers respectively.  {\bf (b)} Comparison between ZT of III-nitride monolayers and that of representative bulk low-temperature TE materials including
     \chem{Bi_{2-x}Sb_{x}Te_{3}} \cite{mao2019high,mao2021thermoelectric},
 \chem{Bi_{2}Te_{3-x}Se_{x}}
 \cite{mao2019high,mao2021thermoelectric},
 \chem{Bi_{0.905}Sb_{0.095}}
\cite{lenoir1996transport,mao2021thermoelectric},
\chem{CePd_{2.95}}
\cite{mahan1997thermoelectric,mao2021thermoelectric},
 \chem{CsBi_{4}Te_{6}}
\cite{chung2000csbi4te6,mao2021thermoelectric},
\chem{Mg_{3}Bi_{1.25}Sb_{0.75}} \cite{pan2020mg,mao2021thermoelectric},
and \chem{YbAl_{3}} \cite{mahan1997thermoelectric,mao2021thermoelectric}.
} \label{fig:lowTZT}
\end{figure*}

The low-temperature ($\leq$300K) thermoelectric figure of merit ZT for both n- and p-type doping of the 2D III-nitride materials was investigated and is presented in Fig.~\ref{fig:lowTZT}. The performance of the monolayers is compared directly with other 2D materials such as TMDs in Fig.~\ref{fig:lowTZT}(a), with the figure of merit at 300K plotted against the bandgap. While all of the semiconducting 2D materials exhibit significant ZT$_{\rm 300K}$, most feature bandgaps substantially smaller than those of the III-nitrides, which at increased temperatures can result in excitation of minority carriers and shift in chemical potential which suppresses the thermoelectric figure of merit. Moreover, unlike TMDs such as \chem{MoS_{2}}, the III-nitride materials are not limited to low temperature TE applications due to onset of sublimation at T $< 500$K \cite{MoS2sub}. In Fig.~\ref{fig:lowTZT}(b), we compare the 2D III-nitrides with established bulk low-temperature TE materials. Under the adopted lattice-conductivity reference, the predicted III-nitride values compare favorably with the selected bulk benchmarks at the sampled low temperatures, with ZT $>2$ obtained for \chem{B_{0.25}Ga_{0.75}N}.

\subsection{Modulation doping and interfacial charge transfer}
\label{sec:modulation-doping}

Wide bandgap semiconductors such as hBN frequently exhibit a doping asymmetry problem (i.e. where only one type of doping -- p-or n-type -- of the material is easy due to band alignment, while the other is difficult). This presents a serious obstacle to realization of wide bandgap semiconductor devices. For wide bandgap nitrides like AlN or hBN, one of the principal difficulties for n-type doping is the close proximity of the CBM to the vacuum level \cite{weston2018native}, which results in a small intrinsic electron affinity $\chi$. The selective alloying of elements that shift the CBM away from the vacuum level increases the electron affinity and dopability of the material \cite{zunger2003practical}. As shown in Fig.~\ref{fig:Figure4}(a), the introduction of group-III elements (Al, Ga) into hBN results in a more favorable band alignment for n-type conduction. In Fig.~\ref{fig:Figure4}(a), band alignment with respect to the vacuum level is shown for the III-nitride monolayers (left).  As with other III-nitrides, the N 2$p$-derived states of the alloy valence band maximum (VBM) are localised and thus exhibit low dispersion (see Fig.~\ref{fig:band}(c) where several flat bands are evident for $\rm x = 0.25, \;0.5, \;0.75$). Materials that would permit work-function-mediated charge transfer -- i.e modulation doping -- of the monolayers \cite{arora2025engineering,chen2022modulation,wang2021modulation,zheng2025ultrafast} are shown on the right of Fig.~\ref{fig:Figure4}(a), and include transition metal oxides (TMOs) and perovskites.

\begin{figure*}[!tp]
    \centering
    \includegraphics[width=0.99\linewidth]{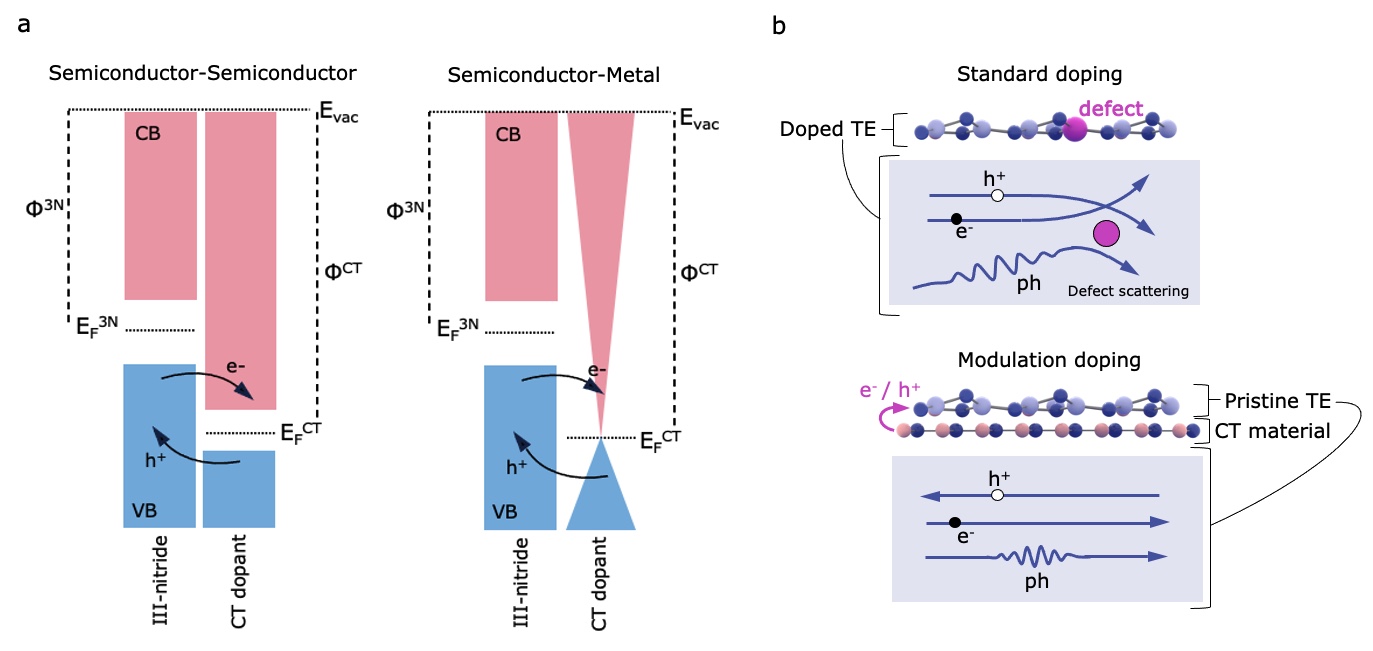}
    \caption{{\bf (a)} Band diagram depicting p-type modulation doping of a 2D III-nitride layer using either another semiconductor that has a broken-gap (i.e type III)  band alignment with $\mathrm{E}^{\mathrm{CT}}_{\mathrm{CBM}} <\mathrm{E}^{\mathrm{III-nitride}}_{\mathrm{VBM}}$  or a metal  with work function $\Phi^{\mathrm{CT}} <\mathrm{E}^{\mathrm{III-nitride}}_{\mathrm{VBM}}$. The electronic bands are modified by the difference in work function between the two materials in the heterostructure and results in transfer of electrons from the 2D III-nitride to the CT material, which hole dopes the 2D III-nitride. {\bf (b)} Comparison between 'standard' substitutional doping in which impurities are introduced into the TE or channel material and modulation doping in which an encapsulation layer donates (accepts)  electrons to (from) the pristine TE or channel material. \cite{chen2022modulation}.
In two-dimensional materials, scattering events from long-range Coulomb interactions are greatly amplified due to reduced dielectric screening in the out-of-plane direction.  This effect can significantly degrade carrier transport and make substitutional doping problematic for atomically thin conduction channels.
\cite{ma2014charge}.
} \label{fig:doping}
\end{figure*}

\begin{figure*}[!tp]
    \centering
    \includegraphics[width=0.78\linewidth]{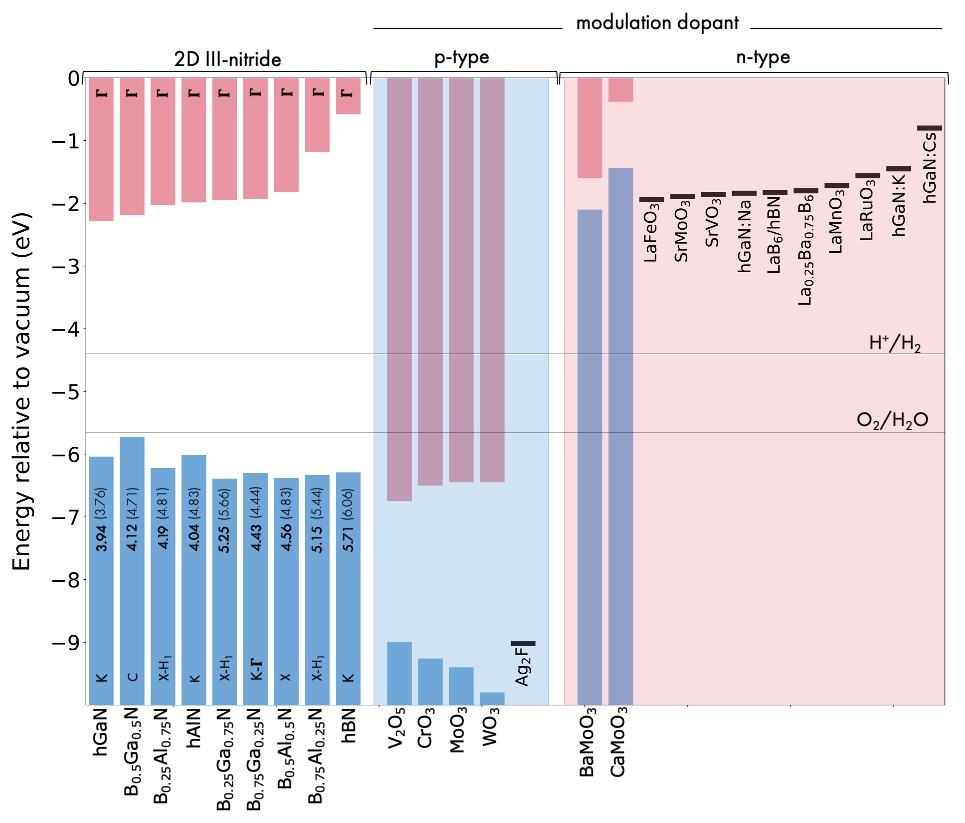}
    \caption{ The band alignment referenced to vacuum level of the valence band maximum (VBM) and conduction band minimum (CBM) energies computed using HSE06. Indirect (direct) bandgap given in units of eV with location in k-space of the conduction (valence) band extrema indicated on each respectively. Materials to engineer interfacial charge-transfer through modulation doping for p- and n-type conduction in the III-nitride monolayers are shown on the right. Proposed p-type modulation dopants include transition metal oxides (TMOs):
\chem{V_2O_3} \cite{laubach2007theoretical,greiner2012universal}, \chem{CrO_3} \cite{zhai2008probing,greiner2012universal},
\chem{MoO_3} \cite{kroger2009p,greiner2012universal},
\chem{WO_3} \cite{meyer2010charge,greiner2012universal}, as well as the metal
\chem{Ag_2F} \cite{schindler2024discovery}; while n-type dopants include perovskites:
\chem{BaMoO_3} \cite{schindler2024discovery,ma2021discovery}, \chem{CaMoO_3} \cite{ma2021discovery},
\chem{LaFeO_3} \cite{jacobs2016understanding,ma2021discovery},
\chem{SrMoO_3} \cite{ma2021discovery},
\chem{SrVO_3} \cite{lin2022demonstration,ma2021discovery,jacobs2016understanding},
\chem{LaMnO_3} \cite{jacobs2016understanding,ma2021discovery},
\chem{LaRuO_3} \cite{jacobs2016understanding},
 alkali-metal-adsorbed hexagonal GaN monolayers: \chem{hGaN: Na, K, Cs} \cite{cui2018alkali},
and hexaborides \chem{LaB_6/hBN} \cite{yamaguchi2023work}, \chem{La_{0.25}Ba_{0.75}B_6} \cite{ma2021work}.
}
    \label{fig:Figure4}
\end{figure*}

Traditional methods of doping such as substitutional doping \cite{suh2014doping}, plasma treatment\cite{chen2013stable}, and ion implantation \cite{nipane2016few}, introduce charge scattering centers which degrade thermoelectric performance. Modulation-doping \cite{bai2026designing,zheng2025ultrafast,xie2024low,wang2021modulation,dingle1978electron,daembkes1991modulation,lee2021remote} can reduce ionized-impurity scattering by spatially separating dopants from the conducting channel, as shown in Fig.~\ref{fig:doping}(b). This separation offers a route to increasing carrier concentration while preserving carrier mobility, with benefits for electronic and thermoelectric transport~\cite{gossard1985modulation}.
Modulation doping has been extensively employed in thin-films, III--V heterostructures, and 2D semiconductors, such as to improve silicon-germanium alloy electrical conductivity by 54\% \cite{yu2012enhancement}, and thermoelectric power factor by 40\% \cite{zebarjadi2011power}. Improvements in $ZT$ have also been demonstrated: Cu-intercalation-mediated modulation doping in n-type \chem{Bi_2Te_{2.7}Se_{0.3}} yielded $ZT=1.42$ at $375~\mathrm K$, approximately $75\%$ above the non-intercalated CuI-doped reference at the same temperature~\cite{chen2022modulation}. These results motivate the material-specific p- and n-type modulation-doping partners proposed here.

For p-type doping, materials with work function deeper than the valence band maximum of the III-nitride monolayers (Ag$_2$F and TMOs) are proposed, which make electron transfer from the valence band of the III-nitrides to the conduction band of the adjacent material favorable and can result in efficient hole doping (see schematic of p-type modulation doping in Fig.  \ref{fig:doping}(a)). Recent works have shown massive interfacial charge transfer in graphene/$\alpha$-\chem{RuCl_{3}} \cite{zhou2019evidence,mashhadi2019spin,chau2024electrical,wang2020modulation,rossi2023direct,rizzo2020charge,vitalone2024charge,balgley2022ultrasharp,qiu2024evidence,rizzo2022nanometer} and WSe$_2$/$\alpha$-\chem{RuCl_{3}} \cite{sternbach2023quenched,rizzo2023polaritonic,xie2024low,pack2024charge} heterostructures with this method.

\begin{figure*}[!tp]
    \centering
    \includegraphics[width=0.99\linewidth]{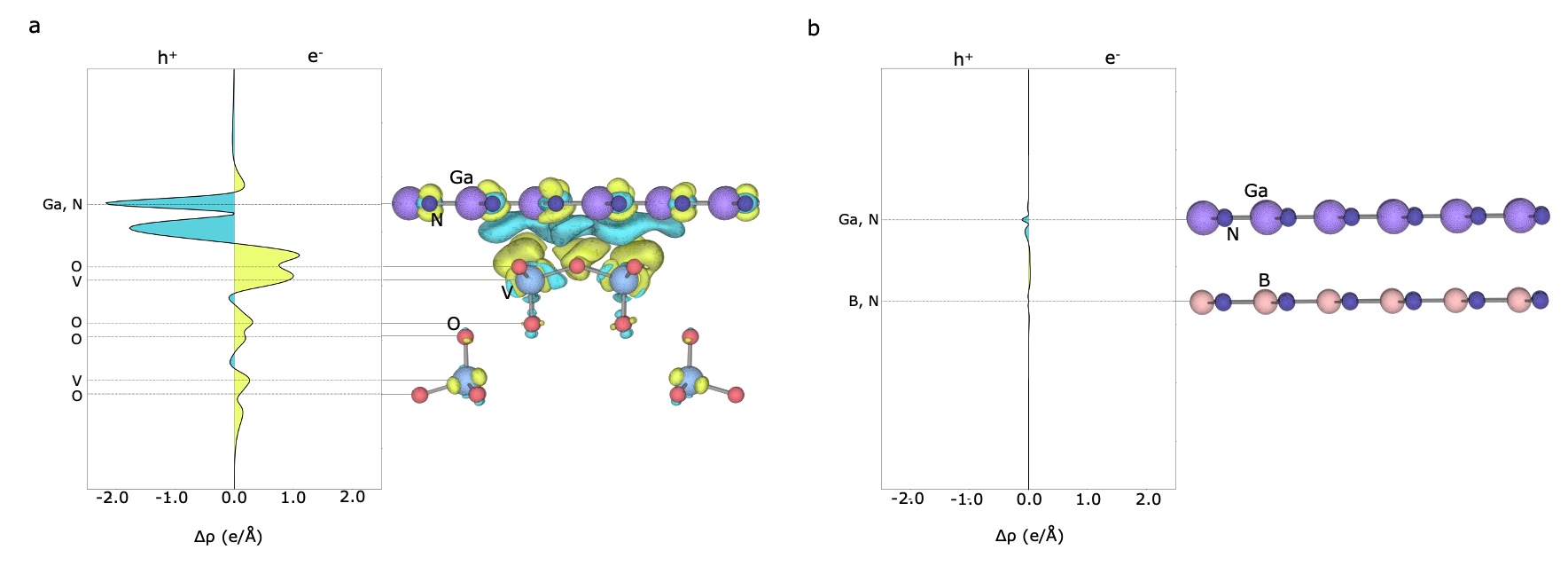}
    \caption{Charge-density difference isosurface of hGaN with {\bf (a)} proposed p-type modulation dopant \chem{V_{2}O_{5}} and {\bf (b)} hBN, which  demonstrates lack of charge transfer between materials without necessary band alignment. The planar-averaged charge-density difference is shown on the left hand side of each subfigure with charge depletion (holes) and accumulation (additional electrons) indicated by blue and yellow respectively. For the hGaN/\chem{V_{2}O_{5}} heterostructure shown in {\bf (a)} each hGaN primitive cell transfers 0.09e to \chem{V_{2}O_{5}}, resulting in a hGaN hole surface charge density of $\sim$1.6$\times10^{14}$cm$^{-2}$. In contrast, an order of magnitude less charge (0.003e) is transferred between layers with unfavorable type-I band alignment, such as demonstrated by the hGaN/hBN heterostructure.
} \label{fig:CDD}
\end{figure*}

Conversely, for n-type doping, materials with work function closer to vacuum level than the III-nitride CBM are proposed, which enable electron transfer from the adjacent material to the III-nitride conduction band. The array of proposed p- and n-type modulation dopants in Fig.~\ref{fig:Figure4} is a distinct contribution of this work: it identifies candidate charge-transfer partners for accessing both hole and electron conduction in the III-nitride monolayers. Figure~\ref{fig:CDD}(a) illustrates the p-type route with calculated charge transfer between hGaN and \chem{V_2O_5}, giving a simulated hole surface charge density of $\sim 1.6\times10^{14}~\mathrm{cm^{-2}}$. Favorable band alignment is used to select the broader candidate set; the explicit hGaN/\chem{V_2O_5} example demonstrates interfacial charge transfer. The literature enhancement in $ZT$ motivates this strategy, while the $ZT$ values reported here are those of the stated monolayer transport model.

\subsection{Electronic properties and transport interpretation}
\label{sec:electronic-response}

\begin{table*}[!tp]
\caption{Electronic properties and estimated intrinsic dielectric breakdown fields of the monolayers. (a) Material-level dielectric response, work function, effective polar-optical-phonon frequency, and estimated breakdown field. (b) Directional band-curvature masses. The valence-band entries $m_{\mathrm v}^*$ retain their negative curvature signs; hole transport is discussed using their magnitudes. Dashes denote values not reported.}
\label{tab:properties}
\centering
\small
\begin{tabular*}{\textwidth}{@{\extracolsep{\fill}}lccccc}
\toprule
\multicolumn{6}{l}{\textbf{(a) Material-level properties}} \\[3pt]
Material & $\varepsilon_{0,\parallel}$ & $\varepsilon_{0,\perp}$ &
\shortstack{$\Phi_{\rm WF}$\\(eV)} &
\shortstack{$\overline{\omega}_{\rm POP}$\\(THz)} &
\shortstack{Estimated $E_{\mathrm b}$\textsuperscript{a}\\(MV/cm)} \\
\midrule
hAlN & 3.54 & 1.83 & 5.07 & \textemdash & 6.23 \\
\chem{B_{0.25}Al_{0.75}N} & 4.09 & 1.76 & 5.40 & 23.98 & 8.90 \\
\chem{B_{0.5}Al_{0.5}N} & 4.42 & 1.73 & 5.36 & 24.29 & 15.95 \\
\chem{B_{0.75}Al_{0.25}N} & 4.11 & 1.85 & 5.30 & 29.25 & 20.06 \\
\addlinespace[3pt]
hGaN & 4.14 & 1.99 & 5.03 & \textemdash & 4.84 \\
\chem{B_{0.25}Ga_{0.75}N} & 3.94 & 2.28 & 5.22 & 28.98 & 10.99 \\
\chem{B_{0.5}Ga_{0.5}N} & 4.87 & 1.72 & 5.25 & 23.42 & 12.98 \\
\chem{B_{0.75}Ga_{0.25}N} & 4.22 & 6.55 & 5.34 & 28.07 & 20.74 \\
\bottomrule
\end{tabular*}
\par\medskip
\begin{tabular*}{\textwidth}{@{\extracolsep{\fill}}lclcl}
\toprule
\multicolumn{5}{l}{\textbf{(b) Directional band-curvature masses}} \\[3pt]
Material & $m_e^*/m_0$ & Direction & $m_{\mathrm v}^*/m_0$ & Direction \\
\midrule
hAlN & 0.52 & $\Gamma \rightarrow \rm{M}$ & -1.47 & $\rm{K} \rightarrow \Gamma$ \\
 & 0.52 & $\Gamma \rightarrow \rm{K}$ & -1.55 & $\rm{K} \rightarrow \rm{M}$ \\
\addlinespace[3pt]
\chem{B_{0.25}Al_{0.75}N} & 0.55 & $\Gamma \rightarrow \rm{X}$ & -1.97 & $\Lambda^{*}\rightarrow \rm{Y}$ \\
 & 0.53 & $\Gamma \rightarrow \rm{Y}$ & \textemdash & \textemdash \\
\addlinespace[3pt]
\chem{B_{0.5}Al_{0.5}N} & 0.68 & $\Gamma \rightarrow \rm{X}$ & -0.93 & $\rm{X} \rightarrow \rm{H}_1$ \\
 & 0.65 & $\Gamma \rightarrow \rm{Y}$ & -4.71 & $\rm{X} \rightarrow \Gamma$ \\
\addlinespace[3pt]
\chem{B_{0.75}Al_{0.25}N} & 0.83 & $\Gamma \rightarrow \rm{M}$ & -0.73 & $\Sigma^{*}\rightarrow \Gamma$ \\
 & \textemdash & \textemdash & -1.3 & $\Sigma^{*} \rightarrow \Gamma$ \\
\addlinespace[3pt]
hGaN & 0.27 & $\Gamma \rightarrow \rm{M}$ & -1.10 & $\rm{K} \rightarrow \Gamma$ \\
 & 0.27 & $\Gamma \rightarrow \rm{K}$ & -1.17 & $\rm{K} \rightarrow \rm{M}$ \\
\addlinespace[3pt]
\chem{B_{0.25}Ga_{0.75}N} & 0.50 & $\Gamma \rightarrow \rm{M}$ & -3.3 & $\Delta^{*}\rightarrow \rm{K}$ \\
 & 0.52 & $\Gamma \rightarrow \rm{K}$ & \textemdash & \textemdash \\
\addlinespace[3pt]
\chem{B_{0.5}Ga_{0.5}N} & 0.54 & $\Gamma \rightarrow \rm{Y}$ & -0.51 & $\rm{C} \rightarrow \rm{H}_1$ \\
 & 0.56 & $\Gamma \rightarrow \rm{X}$ & -0.51 & $\rm{C} \rightarrow \rm{H}$ \\
\addlinespace[3pt]
\chem{B_{0.75}Ga_{0.25}N} & 0.75 & $\Gamma \rightarrow \rm{X}$ & -0.94 & $\Delta^{*}\rightarrow \rm{X}$ \\
 & 0.79 & $\Gamma \rightarrow \rm{Y}$ & \textemdash & \textemdash \\
\bottomrule
\end{tabular*}
\par\smallskip\noindent
\begin{minipage}{\textwidth}
\footnotesize
\raggedright
\textsuperscript{a} Kim--Pilania--Ramprasad intrinsic-breakdown
estimates~\cite{kim2016breakdown}, using $E_{\mathrm g}$ and the maximum
phonon frequency $\nu_{\max}$, distinct from the weighted effective
polar-optical frequency $\overline{\omega}_{\rm POP}$ listed here.
Numerical outputs are rounded to two decimal places; the equation,
maximum-frequency inputs, and composition trend are in the Supplemental
Material~\cite{supplemental}, Sec.~SV, Table~S1, and Fig.~S3.
\par\smallskip
Low-symmetry point coordinates: $\Lambda$: $[0.31,0.35,0.00]$;
$\Sigma$: $[0.66,-0.35,0.00]$; $\Delta$: $[0.04,0.04,0.00]$.
\end{minipage}
\end{table*}

Table~\ref{tab:properties} connects the composition-dependent electronic
structure in Fig.~\ref{fig:band} to the transport and charge-transfer
properties discussed above. Band curvature sets the carrier velocities and
contributes to the density of states, dielectric polarization screens
electrostatic perturbations, and the work function describes the Fermi-level
position relative to vacuum. The effective optical-phonon frequency provides
an energy scale for an additional carrier-scattering channel. These
quantities therefore describe different contributions to the electronic
response rather than interchangeable measures of thermoelectric performance.

The electron masses increase with boron content over the compositions
reported in the table. In the Al-containing series, the values increase
from $0.52m_0$ in hAlN to $0.53$--$0.55m_0$ at $x=0.25$,
$0.65$--$0.68m_0$ at $x=0.5$, and $0.83m_0$ at $x=0.75$.
The corresponding Ga-containing values increase from $0.27m_0$ in hGaN
to $0.50$--$0.52m_0$, $0.54$--$0.56m_0$, and
$0.75$--$0.79m_0$. The similar masses along the two tabulated directions
at $x=0.25$ and $0.5$ indicate relatively weak directional variation of
the local conduction-band curvature. In a parabolic-band picture,
$\mu_\alpha=e\tau_\alpha/m_{\mathrm c,\alpha}$, where
$m_{\mathrm c,\alpha}$ is the conductivity mass. Thus, the increasing
electron mass would lower mobility at a fixed relaxation time. The
conductivity comparison between the Al- and Ga-containing alloys cannot,
however, be inferred from these masses alone: the different conductivities
in Fig.~S1 of the Supplemental Material~\cite{supplemental} also reflect
their material-dependent MRTA lifetimes and the full occupied band
dispersion.

The hole masses show a more pronounced and nonmonotonic composition
dependence. At $x=0.25$, the reported magnitudes are $1.97m_0$ for
\chem{B_{0.25}Al_{0.75}N} and $3.3m_0$ for
\chem{B_{0.25}Ga_{0.75}N}, larger than the corresponding electron
masses and the hole masses of hAlN ($1.47$--$1.55m_0$) and hGaN
($1.10$--$1.17m_0$). These values quantify the flattening of the
N-$p$-derived valence-band edge associated with puckering and orbital
mixing in Fig.~\ref{fig:band}. Within a parabolic-band interpretation,
a larger density-of-states mass can increase $|S|$ at a fixed sheet
carrier density, while a larger mass along the transport direction reduces
carrier velocities. This competition helps interpret the generally larger
thermopower and smaller conductivity for p-type than for n-type transport
at $x=0.25$ in Fig.~S1. At $x=0.5$, the reported hole masses of the
Al-containing alloy, $0.93m_0$ and $4.71m_0$, differ by approximately a
factor of five, whereas the two reported Ga-containing hole masses are
both $0.51m_0$. Consequently, a uniformly heavy or strongly anisotropic
hole band is not a property of every alloy in these series.

The directional contrast in \chem{B_{0.5}Al_{0.5}N} illustrates why
density-of-states and conductivity masses must be distinguished. A heavy
direction can supply a large density of states while a lighter direction
supports faster transport, a combination that can improve the balance
between thermopower and conductivity~\cite{parker2015anisotropy}.
The tabulated line curvatures do not establish the complete mass tensor,
valley multiplicity, or energy separation between pockets. These factors,
and the increase in scattering phase space that can accompany a larger
density of states, determine whether anisotropy improves the power
factor~\cite{park2021optimal}. In particular, the large hole mass of
\chem{B_{0.25}Ga_{0.75}N} pertains to its p-type response; the strong
n-type performance in Fig.~\ref{fig:lowTZT} involves its more dispersive
conduction band, with masses of $0.50$--$0.52m_0$, together with the
calculated electron lifetimes.

The static dielectric constants reveal a different composition dependence.
The in-plane values lie between $3.54$ and $4.87$, while the out-of-plane
values are $1.72$--$2.28$ for all entries except
\chem{B_{0.75}Ga_{0.25}N}. For this alloy,
$\varepsilon_{0,\perp}=6.55$, exceeding its in-plane value of $4.22$
and reversing the dielectric anisotropy of the other tabulated materials.
This enhanced perpendicular response suggests stronger screening of
slowly varying fields normal to the layer, relevant to remote charges and
interfacial charge transfer. Static dielectric screening also enters the
piezoelectric-scattering model, together with the piezoelectric and elastic
tensors~\cite{ganose2021scattering}; a larger individual dielectric
component alone therefore does not establish a longer lifetime or higher
conductivity. The tabulated permittivities include electronic and ionic
polarization and use the effective-thickness treatment described in the
Supplemental Material~\cite{laturia2018dielectric,supplemental}.

The estimated intrinsic breakdown fields in Table~\ref{tab:properties}(a)
rise from $6.23$ (hAlN) and $4.84$ (hGaN) $\mathrm{MV/cm}$ to $20.06$
and $20.74~\mathrm{MV/cm}$ at $x=0.75$, respectively. These B-rich
values indicate high predicted intrinsic field tolerance of interest for
power electronics. For scale, a literature assessment gives bulk GaN and
diamond critical fields of approximately $3.4$ and $10~\mathrm{MV/cm}$,
respectively, and identifies the commonly quoted $8~\mathrm{MV/cm}$ for
$\beta$-\chem{Ga_2O_3} as an estimate~\cite{slobodyan2022critical}.
Differences in dimensionality, field profiles, and methods make these
comparisons indicative. The present values are screening estimates from the
Kim--Pilania--Ramprasad model~\cite{kim2016breakdown}, using the band gap
and maximum phonon frequency $\nu_{\max}$; their application to monolayers
requires validation. The equation and full trend are given in Sec.~SV and
Fig.~S3 of the Supplemental Material~\cite{supplemental}.

The work functions span $5.03$--$5.40$~eV and show that the vacuum-referenced
electronic energies are tunable by alloying. Relative to hAlN ($5.07$~eV),
the Al-containing alloys have higher work functions, reaching $5.40$~eV
at $x=0.25$ before decreasing to $5.30$~eV at $x=0.75$.
The Ga-containing values increase from $5.03$~eV in hGaN to $5.22$,
$5.25$, and $5.34$~eV with increasing $x$. Since
$\Phi_{\rm WF}=E_{\rm vac}-E_{\rm F}$, these changes describe the
Fermi-level offsets relevant to equilibration with a contacting material.
For a semiconductor, the work function also depends on Fermi-level
placement and does not specify the electron affinity or ionization energy
by itself. The vacuum-referenced band edges in Fig.~\ref{fig:Figure4}
are therefore needed to select electron-donating and electron-accepting
partners. The work-function trends complement that selection, while the
explicit interface calculation in Fig.~\ref{fig:CDD} accounts for charge
redistribution in the hGaN/\chem{V_2O_5} example.

The effective polar-optical-phonon frequencies range from $23.42$ to
$29.25$~THz for the six mixed alloys. They increase with $x$ in the
Al-containing series, whereas the Ga-containing series has its lowest
value at $x=0.5$. At a fixed temperature, a higher phonon energy reduces
the Bose occupation relevant to absorption and raises the carrier-energy
threshold for emission. This motivates examining reduced polar-optical
scattering, but the frequency alone cannot establish that this mechanism
is negligible. Its strength also depends on the coupling matrix elements,
electronic and ionic screening, and the dielectric
environment~\cite{sohier2016frohlich,sohier2017erratum}.
The reported transport values retain the ADP-plus-PIE approximation in
Sec.~\ref{sec:scattering-model}. Moreover, a weighted zone-center optical
frequency does not determine lattice thermal conductivity; no
composition-dependent $\kappa_{\mathrm l}$ is inferred from this column
of Table~\ref{tab:properties}.

Together, these results explain how alloying changes the ingredients of
the thermoelectric response. The density of states, carrier velocities,
and screened scattering jointly determine the energy-dependent transport
distribution, and hence $\sigma$, $S$, and $\kappa_{\mathrm e}$.
Reduced dimensionality further changes the density-of-states profile,
providing the quantum-confinement context illustrated in Fig.~S2 of the
Supplemental Material~\cite{hicks1993effect,hicks1993thermoelectric,mao2016size,supplemental}.
The sensitivity of thermopower to these ingredients can be illustrated
by the Mott relation in the degenerate limit~\cite{nye1985physical},
\begin{equation}
S_\alpha\simeq-\frac{\pi^2k_{\mathrm B}^2T}{3e}
\left.\frac{\partial\ln\Xi_\alpha(E;T,\mu)}{\partial E}\right|_{E=E_{\mathrm F}},
\label{eq:mott}
\end{equation}
where $e>0$ and $\Xi_\alpha$ is the transport distribution defined in
Sec.~\ref{sec:scattering-model}. Its logarithmic derivative contains both
band-dispersion and scattering contributions. The finite-temperature
coefficients are obtained from the full transport moments, which retain
the state-dependent lifetimes. Accordingly, the large $ZT$ values in
Figs.~\ref{fig:ZT} and~\ref{fig:lowTZT} reflect a balance between the
power factor $S^2\sigma$ and electronic heat conduction, evaluated with
the common lattice-conductivity reference. Table~\ref{tab:properties}
provides a physical interpretation of that balance, while the full
transport calculation is needed to establish the ranking of compositions
and carrier types.

\section{Computational methods}
\label{sec:methods}
\subsection{Carrier scattering and electronic transport}
\label{sec:scattering-model}
\begin{figure*}[!tp]
    \centering
    \includegraphics[width=0.99\textwidth,height=0.28\textheight,keepaspectratio]{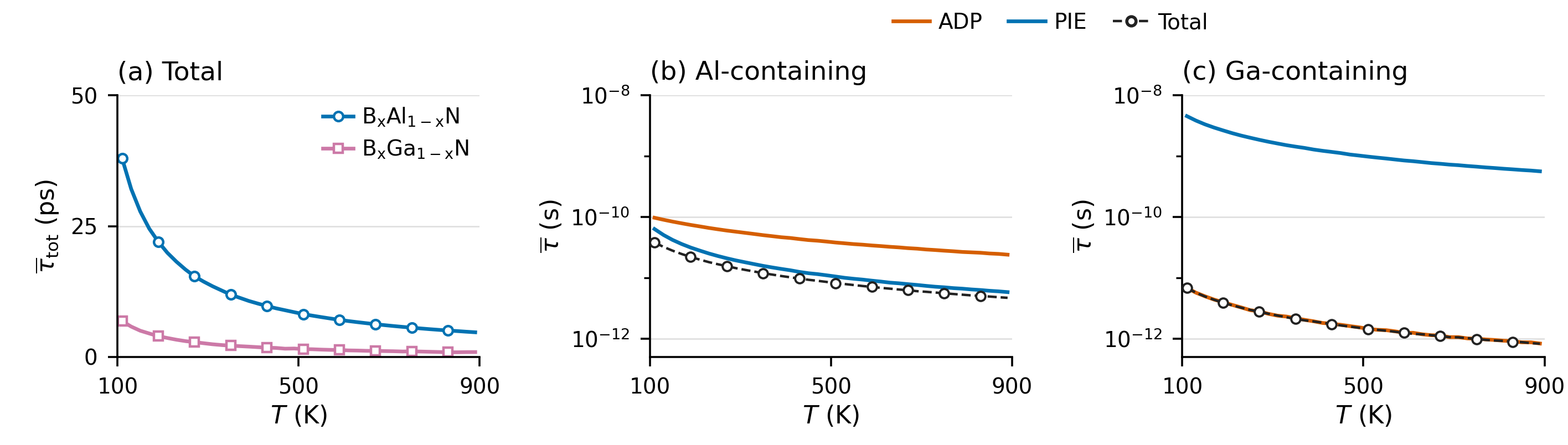}
    \caption[Temperature-dependent average MRTA relaxation times]{Temperature-dependent average MRTA relaxation times. 
    (a) Total averages for \chem{B_xAl_{1-x}N} and \chem{B_xGa_{1-x}N}.
    (b),(c) Acoustic deformation-potential (ADP), piezoelectric (PIE), and total averages for the Al- and Ga-containing alloys, respectively.}
    \label{fig:scattering}
\end{figure*}

Electronic structure and semiclassical transport were investigated using
density functional theory, BoltzTraP2~\cite{madsen2018boltztrap2}, and
AMSET~\cite{ganose2021scattering}. Carrier scattering was treated within the
momentum relaxation-time approximation (MRTA), with a momentum relaxation
time $\tau_{n\mathbf{k}}(T,\mu)$ evaluated for each electronic band $n$ and
wave vector $\mathbf{k}$ at temperature $T$ and chemical potential $\mu$.
The MRTA accounts for the change in carrier velocity associated with a
scattering event when evaluating momentum relaxation. The state-dependent
lifetimes were retained in the transport calculation rather than replaced
by a common constant.

The scattering model includes acoustic deformation-potential (ADP) and
piezoelectric (PIE) scattering. ADP scattering describes the coupling between
carriers and the band-energy shifts induced by acoustic lattice
deformations. PIE scattering arises from the electric fields generated by
piezoelectric polarization accompanying these deformations. Scattering
probabilities were evaluated within the Born approximation using AMSET,
with first-principles deformation potentials, elastic properties,
piezoelectric coefficients, and dielectric response as material-specific
inputs~\cite{ganose2021scattering}. The contributions of the included
mechanisms were combined for each electronic state according to
Matthiessen's rule,
\begin{equation}
\frac{1}{\tau_{n\mathbf{k}}(T,\mu)}
=\Gamma^{\mathrm{ADP}}_{n\mathbf{k}}(T,\mu)
+\Gamma^{\mathrm{PIE}}_{n\mathbf{k}}(T,\mu),
\label{eq:scattering-total}
\end{equation}
where $\Gamma^i_{n\mathbf{k}}$ is the momentum relaxation rate associated
with scattering mechanism $i$.

Only in-plane electronic transport is considered. Let
$\Xi_\alpha(E;T,\mu)$ denote the volumetric transport distribution along
an in-plane principal direction $\alpha$. It includes the band velocities,
electronic-state weights, and state-dependent MRTA lifetimes. With
$f(E;T,\mu)$ the Fermi--Dirac distribution, the transport moments are
\begin{equation}
L_\alpha^{(m)}=\int dE\,\Xi_\alpha(E;T,\mu)(E-\mu)^m
\left(-\frac{\partial f}{\partial E}\right).
\label{eq:transport-moments}
\end{equation}
For electron charge $-e$, with $e>0$, the principal-direction transport
coefficients are
\begin{align}
\sigma_\alpha&=e^2L_\alpha^{(0)},\label{eq:transport-sigma}\\
S_\alpha&=-\frac{L_\alpha^{(1)}}{eT L_\alpha^{(0)}},\label{eq:transport-seebeck}\\
\kappa_{\mathrm e,\alpha}&=\frac{1}{T}\left[
L_\alpha^{(2)}-\frac{(L_\alpha^{(1)})^2}{L_\alpha^{(0)}}\right].
\label{eq:transport-kappae}
\end{align}
Here, $\kappa_{\mathrm e,\alpha}$ is the electronic thermal conductivity at
zero electrical current. A lifetime common to all electronic states
factors out of every moment and cancels from $S_\alpha$. State-dependent
lifetimes generally do not cancel; the present calculation retains the
AMSET MRTA lifetimes in the transport response~\cite{madsen2018boltztrap2,ganose2021scattering}.
The temperature-dependent averages in Fig.~\ref{fig:scattering} summarize
these lifetimes and are not substituted for them in the transport integrals.

Calculations were performed at multiple temperatures and carrier
concentrations within the rigid-band approximation: the electronic bands
are held fixed while occupations change with temperature and chemical
potential. The relation between volumetric and sheet transport quantities
is defined in Sec.~\ref{sec:normalization}. Additional computational
parameters are given in the Supplemental Material~\cite{supplemental}.

Polar-optical-phonon scattering was omitted from the reported calculations.
This approximation was motivated by the high effective optical-phonon
frequencies calculated for the mixed nitride monolayers, approximately
$20$--$30$~THz. Higher phonon energies reduce thermal phonon populations,
while the magnitude of polar-optical scattering also depends on coupling
strength and the availability of absorption and emission processes.
In two-dimensional materials, the Fr{\"o}hlich interaction depends on
dielectric screening and the surrounding
environment~\cite{sohier2016frohlich,sohier2017erratum}.
The resulting transport predictions therefore apply to the stated
ADP-plus-PIE scattering model.

\subsection{Effective polar-optical-phonon frequency}
The effective polar phonon frequency $\overline{\omega}_{\mathrm{POP}}$ was determined from the phonon frequencies and eigenvectors generated by DFPT in VASP. Each $\Gamma$-point phonon mode from the DFPT run was weighted by the associated dipole moment as
\begin{equation}
w_\nu=\sum_\kappa\left[\frac{1}{M_\kappa \omega_{\mathbf{q} \nu}}\right]^{1 / 2} \times\left[\mathbf{q} \cdot \mathbf{Z}_\kappa^* \cdot \mathbf{e}_{\kappa \nu}(\mathbf{q})\right]
\end{equation}
where $\mathbf{Z}_\kappa^*$ is the Born effective charge, and then averaged over the unit sphere to capture dipole anisotropy and polar divergence as $\rm q \rightarrow 0$.  The effective phonon frequency is then given by the weighted sum of $\Gamma$-point phonon modes
\begin{equation}
\overline{\omega}_{\mathrm{POP}}=\frac{\omega_{\Gamma \nu} w_\nu}{\sum_\nu w_\nu}.
\end{equation}

\subsection{Lattice heat transport: literature-based reference}
\label{sec:lattice-model}
We represent lattice heat transport using the common reference
$\kappa_{\mathrm l}^{\mathrm{ref}}=550~\mathrm{W\,m^{-1}\,K^{-1}}$,
held fixed across all compositions and sampled temperatures. This
literature-motivated reference replaces an alloy-specific phonon-transport
calculation. It is motivated by the high room-temperature in-plane
conductivity discussed for bulk hBN~\cite{yuan2019hbnthermal}. For
comparison, first-principles calculations report
$14.93~\mathrm{W\,m^{-1}\,K^{-1}}$ for planar monolayer
GaN~\cite{qin2017ganthermal} and
$306.5~\mathrm{W\,m^{-1}\,K^{-1}}$ for unstrained monolayer AlN at
$300~\mathrm K$~\cite{banerjee2022alnthermal}. Within a common normalization,
using the higher reference gives smaller $ZT$ estimates than using either
of these lower lattice conductivities with the same electronic
transport coefficients.

For fixed $S$, $\sigma$, and $\kappa_{\mathrm e}$, the reference estimate is
a lower bound with respect to the lattice contribution whenever
\begin{equation}
\kappa_{\mathrm l}(T)\leq\kappa_{\mathrm l}^{\mathrm{ref}}
\quad\Longrightarrow\quad
ZT(T)\geq ZT^{\mathrm{ref}}(T).
\label{eq:conditional-bound}
\end{equation}
This comparison requires consistent normalization of all conductivities.
Alloying and chemical ordering can modify phonon transport
nonmonotonically~\cite{wang2021alloythermal}. The cited room-temperature
values therefore motivate a high-conductivity reference scenario without
establishing an upper bound for every alloy and temperature. Both the
low-temperature results and the wider temperature survey use this same
fixed lattice reference. The electronic transport coefficients and
carrier relaxation times retain their calculated temperature and
carrier-concentration dependence.

\subsection{Two-dimensional normalization}
\label{sec:normalization}
For a monolayer, we distinguish the periodic supercell height $c$ normal
to the layer from the effective material thickness $t$. Following the
relaxed-bilayer thickness convention of Laturia
et al.~\cite{laturia2018dielectric}, $t$ is defined by the interlayer
separation in the relaxed bilayer. For the six mixed alloys
($x=0.25$, $0.5$, and $0.75$ in the Al- and Ga-containing series), this
value is $7$~\AA{}. For in-plane cell area $A$,
supercell-normalized coefficients refer to the volume $Ac$,
whereas effective volumetric coefficients refer to $At$. The corresponding
sheet electrical and electronic thermal conductances are
\begin{align}
G_{\square,\alpha}
&=c\,\sigma_{\mathrm{SC},\alpha}=t\,\sigma_\alpha,
\label{eq:sheet-electrical}\\
K_{\square,\mathrm e,\alpha}
&=c\,\kappa_{\mathrm e,\mathrm{SC},\alpha}
=t\,\kappa_{\mathrm e,\alpha}.
\label{eq:sheet-electronic-thermal}
\end{align}
Thus, conversion from the supercell convention to an effective material
thickness multiplies both electronic conductivities by $c/t$. The Seebeck
coefficient is unchanged by this common rescaling. Sheet conductances
provide a thickness-independent description of monolayer
transport~\cite{wu2017sheet}; $G_\square$ has units of S and $K_\square$
has units of $\mathrm{W\,K^{-1}}$.

The sheet carrier concentration is
\begin{equation}
n_{\mathrm{2D}}=c\,n_{\mathrm{SC}}=t\,n,
\label{eq:sheet-density}
\end{equation}
where $n_{\mathrm{SC}}$ and $n$ are carrier concentrations normalized to
$Ac$ and $At$, respectively. Length and concentration units must be
compatible; for example, $c$ in cm and $n_{\mathrm{SC}}$ in
$\mathrm{cm^{-3}}$ give $n_{\mathrm{2D}}$ in $\mathrm{cm^{-2}}$.

Defining the lattice sheet thermal conductance by
$K_{\square,\mathrm l,\alpha}=t\,\kappa_{\mathrm l,\alpha}$, the
principal-direction figure of merit is
\begin{equation}
ZT_\alpha=\frac{S_\alpha^2G_{\square,\alpha}T}
{K_{\square,\mathrm e,\alpha}+K_{\square,\mathrm l,\alpha}}.
\label{eq:sheet-zt}
\end{equation}
For consistency with electronic coefficients expressed using thickness
$t$, the lattice reference must be represented by
$K_{\square,\mathrm l}^{\mathrm{ref}}=t\,\kappa_{\mathrm l}^{\mathrm{ref}}$.
Equation~\ref{eq:sheet-zt} is invariant under a change in $t$ when the
physical sheet conductances are held fixed and all volumetric
conductivities are rescaled consistently. Holding the numerical value
$550~\mathrm{W\,m^{-1}\,K^{-1}}$ fixed while changing $t$ instead changes
the reference sheet heat conductance and defines a different model.

\section{Summary and conclusions}

We have investigated the electronic structure, thermoelectric response,
and estimated intrinsic breakdown fields of ordered monolayer
\chem{B_xAl_{1-x}N} and \chem{B_xGa_{1-x}N}.
Composition-induced puckering changes the band-edge dispersion and
anisotropy. The band-curvature masses, dielectric response, work functions,
and optical-phonon descriptors connect these changes to carrier transport
and interfacial charge transfer. The p- and n-type responses reflect their
respective valence and conduction bands together with state-dependent
carrier lifetimes.

Within the acoustic deformation-potential plus piezoelectric scattering
model, selected p- and n-type monolayers have predicted maximum $ZT>2.5$;
n-type \chem{B_{0.25}Ga_{0.75}N} reaches $ZT>2$ at $300~\mathrm K$.
These results motivate further investigation of low-temperature
thermoelectric applications. All estimates use the same
temperature-independent lattice reference,
$550~\mathrm{W\,m^{-1}\,K^{-1}}$, rather than alloy-specific phonon
transport. With electronic coefficients held fixed and conductivities
normalized consistently, this gives a lower bound with respect to the
lattice contribution only when the actual lattice conductivity does not
exceed the reference. The omission of polar-optical scattering and other
unmodeled mechanisms remains a limitation of the quantitative transport
predictions.

Vacuum-referenced band alignments identify candidate modulation-doping
partners for both carrier types. The explicit hGaN/\chem{V_2O_5}
calculation yields a hole surface density of approximately
$1.6\times10^{14}~\mathrm{cm^{-2}}$, illustrating the p-type charge-transfer
route. Transport in these heterostructures remains to be evaluated to
determine how the spatial separation of dopants and carriers affects
attainable thermoelectric performance.

The Kim--Pilania--Ramprasad estimates increase from $6.23$ and
$4.84~\mathrm{MV/cm}$ in hAlN and hGaN to $20.06$ and
$20.74~\mathrm{MV/cm}$ in their respective $x=0.75$ alloys.
These high predicted intrinsic fields motivate investigation of the
B-rich alloys for power electronics alongside their thermoelectric
response. Their assessment uses the band gap and maximum phonon frequency,
distinct from the effective polar-optical-phonon frequency. Validation for
monolayers requires attention to field direction, contacts, defects, and
device geometry; the scalar estimates do not establish a device
breakdown voltage.

The tellurium-free chemistry and preliminary constituent-cost comparison
provide additional motivation, while fabrication costs and alloy-specific
biocompatibility remain to be established. Alloy-resolved lattice heat
transport, additional carrier-scattering mechanisms, doped-interface
transport, and direction-resolved breakdown studies are needed to assess
the experimentally attainable thermoelectric and high-field performance.

\nocite{madsen2018boltztrap2,ganose2021scattering,kresse1996efficiency,farris2018theory,ridley1988quantum,ridley1998review,kim2016breakdown,hattori2016anisotropic}
\bibliographystyle{apsrev4-2}
\bibliography{references}
\end{document}


\title{Supplemental Material for ``Thermoelectric properties and modulation doping of ultrawide-bandgap nitride monolayers''}
\author{Ariel R. Barr}
\email{arbarr@mit.edu}
\affiliation{Department of Materials Science \& Engineering, Massachusetts Institute of Technology, Cambridge, Massachusetts 02139, USA}
\renewcommand{\thesection}{S\Roman{section}}
\renewcommand{\thefigure}{S\arabic{figure}}
\renewcommand{\thetable}{S\arabic{table}}
\renewcommand{\theequation}{S\arabic{equation}}
\maketitle

\section{Computational methods}
\label{sec:si-computational-methods}
\setcounter{figure}{0}

The Boltzmann transport coefficients were computed within the momentum relaxation-time approximation (MRTA), using state-dependent carrier scattering rates calculated with AMSET. BoltzTraP2 (BT2) \cite{madsen2018boltztrap2} and AMSET~\cite{ganose2021scattering} were used for the electronic-structure-based transport analysis. In this framework, relaxation times are evaluated for individual band and wave-vector states rather than imposed as a common constant. Uniform sampling of the monolayer electronic bands was performed using  VASP \cite{kresse1996efficiency} on a $31\times 31 \times 3$ $\Gamma$-centered grid with 15~\AA{} vacuum, and interpolated using BT2 under a rigid band approximation (i.e. the bands do not change with doping or temperature) with k-point sampling 30 times more dense than the ab initio mesh, which was approximately equivalent to a k-point mesh of $53\times53\times17$. \\

For each band $n$ and wave vector $\mathbf{k}$, AMSET yields a momentum
relaxation time $\tau_{n\mathbf{k}}(T,\mu)$. The transport distribution
$\Xi_\alpha(E;T,\mu)$ retains these state-dependent lifetimes together with
band velocities and electronic-state weights. The main article defines the
transport moments and the resulting $\sigma_\alpha$, $S_\alpha$, and
zero-current electronic thermal conductivity $\kappa_{\mathrm e,\alpha}$.
A lifetime common to all states cancels from the Seebeck coefficient;
state-dependent lifetimes generally do not~\cite{ganose2021scattering,farris2018theory}.
The temperature-dependent average lifetimes shown in the main article
summarize the MRTA calculation and are not used as common lifetimes in
place of the state-resolved values. Only in-plane transport is considered.

State- and temperature-dependent scattering rates were evaluated within
the Born approximation using AMSET~\cite{ganose2021scattering}. Inputs
include contraction- and expansion-averaged acoustic deformation
potentials, elastic properties, and the piezoelectric tensor obtained
using density functional perturbation theory (DFPT) in VASP. The
deformation-potential tensor describes band-energy shifts under
homogeneous strain,
\begin{equation}
D_{n\mathbf{k},\alpha\beta}
=\frac{\partial\varepsilon_{n\mathbf{k}}}{\partial\epsilon_{\alpha\beta}},
\end{equation}
where $\epsilon_{\alpha\beta}$ is the strain tensor.

For the included ADP and PIE mechanisms, the state-dependent momentum
relaxation rates are combined using Matthiessen's rule,
\begin{equation}
\frac{1}{\tau_{n\mathbf{k}}(T,\mu)}
=\Gamma^{\mathrm{ADP}}_{n\mathbf{k}}(T,\mu)
+\Gamma^{\mathrm{PIE}}_{n\mathbf{k}}(T,\mu).
\end{equation}
Here, $\Gamma^i_{n\mathbf{k}}=1/\tau^i_{n\mathbf{k}}$ denotes the momentum
relaxation rate for mechanism $i$~\cite{farris2018theory,ridley1988quantum,ridley1998review}.
Polar-optical-phonon scattering was omitted from the reported transport
calculation, as described in the main article. The effective optical-phonon
frequencies are approximately $20$--$30$~THz for the mixed alloys, compared
with approximately $8$~THz in GaAs. Related studies provide context for the
high-frequency argument~\cite{liu2018simultaneously,zhang2023electron,bhattacharjee2017nitride}.

The two-dimensional Fr\"ohlich interaction also depends on dimensionality,
dielectric screening, and the surrounding environment~\cite{sohier2016frohlich,sohier2017erratum}.
The omission of polar-optical scattering is therefore an approximation of the
present transport model, whose applicability depends on these additional
coupling and screening effects.

The effective polar phonon frequency $\omega_{\mathrm{eff}}$ was determined from the phonon frequencies and eigenvectors generated by DFPT in VASP. Each $\Gamma$-point phonon mode from the DFPT run was weighted by the associated dipole moment as
\begin{equation}
w_\nu=\sum_\kappa\left[\frac{1}{M_\kappa \omega_{\mathbf{q} \nu}}\right]^{1 / 2} \times\left[\mathbf{q} \cdot \mathbf{Z}_\kappa^* \cdot \mathbf{e}_{\kappa \nu}(\mathbf{q})\right]
\end{equation}
where $\mathbf{Z}_\kappa^*$ is the Born effective charge, and then averaged over the unit sphere to capture dipole anisotropy and polar divergence as $\rm q \rightarrow 0$.  The effective phonon frequency is then given by the weighted sum of $\Gamma$-point phonon modes
\begin{equation}
\omega_{\mathrm{eff}}=\frac{\omega_{\Gamma \nu} w_\nu}{\sum_\nu w_\nu}.
\end{equation}

The static dielectric response was calculated using the method of
Laturia et al.~\cite{laturia2018dielectric}, with $20$~\AA{} vacuum in the
out-of-plane direction. Both electronic and ionic contributions are
included in the static dielectric constants. The supercell dielectric
response contains a vacuum contribution, which is removed using the
out-of-plane and in-plane rescaling relations given as Eqs.~(1) and~(2)
in Ref.~\cite{laturia2018dielectric}:
\begin{align}
\varepsilon_{\mathrm{2D},\perp}
&=\left[1+\frac{c}{t}\left(
\frac{1}{\varepsilon_{\mathrm{SC},\perp}}-1\right)\right]^{-1},
\label{eq:dielectric-perp}\\
\varepsilon_{\mathrm{2D},\parallel}
&=1+\frac{c}{t}\left(\varepsilon_{\mathrm{SC},\parallel}-1\right).
\label{eq:dielectric-parallel}
\end{align}
Here, $c$ is the complete supercell height, including the vacuum, and
$t$ is the effective monolayer thickness. Following the relaxed-bilayer
convention, $t$ is obtained from the interlayer separation in the relaxed
bilayer, rather than the vertical extent of one puckered monolayer. The
reported value is $7$~\AA{} for each of the six mixed alloys
(\chem{B_xM_{1-x}N}, with $M=\mathrm{Al},\mathrm{Ga}$ and
$x=0.25$, $0.5$, $0.75$). The in-plane dielectric response is reported as
an in-plane average. These equations rescale dielectric constants;
the conversion of electronic transport coefficients is defined separately
in Sec.~\ref{sec:si-normalization}.

\begin{figure}
    \centering
    \includegraphics[width=0.99\linewidth]{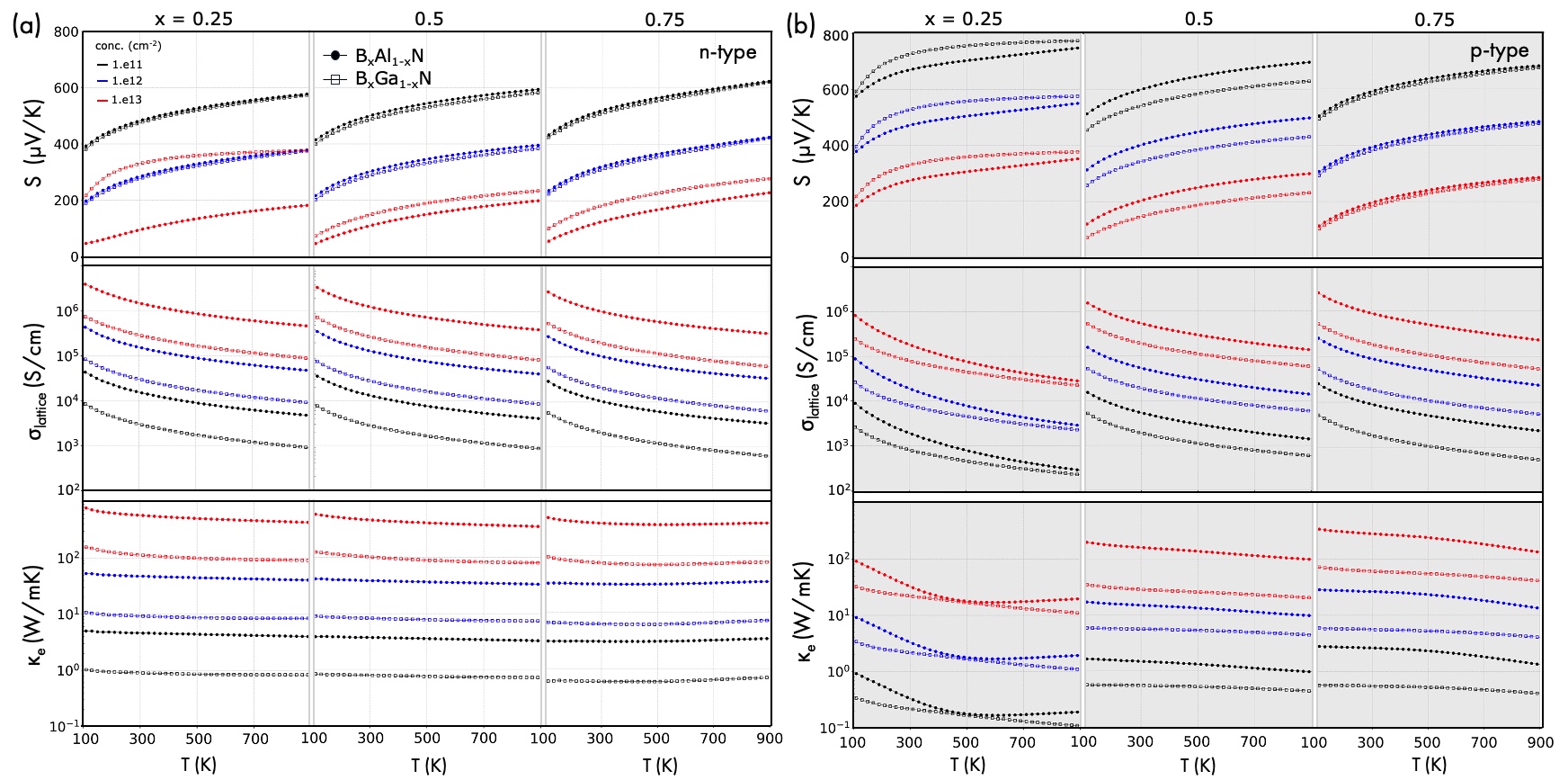}
    \caption{Temperature-dependent transport properties of \chem{B_xAl_{1-x}N} and \chem{B_xGa_{1-x}N} monolayers as a function of {\bf (a)} n-type and {\bf (b)} p-type surface carrier concentration. The Seebeck coefficient and electrical conductivity are given as the magnitude of the in-plane tensor elements (i.e. $|| (\rm{S_{xx}+S_{yy}})/2||$))).
}
\label{fig:transport}
\end{figure}

\section{Lattice heat transport: literature-based reference}
The $ZT$ estimates use the common lattice reference
$\kappa_{\mathrm l}^{\mathrm{ref}}=550~\mathrm{W\,m^{-1}\,K^{-1}}$
at every sampled composition and temperature. Its literature rationale
and conditional lower-bound interpretation are described in the main
article: bulk hBN motivates the high reference~\cite{yuan2019hbnthermal},
with lower published monolayer GaN and AlN values providing
context~\cite{qin2017ganthermal,banerjee2022alnthermal}. No alloy-specific
lattice-conductivity curve was calculated. For a fixed common
normalization, the reference figure of merit is
\begin{equation}
ZT^{\mathrm{ref}}(T,n_{\mathrm{2D}})=
\frac{\sigma(T,n_{\mathrm{2D}})S^2(T,n_{\mathrm{2D}})T}
{\kappa_{\mathrm e}(T,n_{\mathrm{2D}})+\kappa_{\mathrm l}^{\mathrm{ref}}}.
\end{equation}
The fixed lattice reference is separate from the temperature-dependent
MRTA lifetimes and electronic coefficients. At fixed $S$, $\sigma$, and
$\kappa_{\mathrm e}$, it gives a lower bound with respect to the lattice
contribution only when the actual $\kappa_{\mathrm l}(T)$ does not exceed
the reference in the same normalization. The room-temperature literature
values do not establish this inequality for every alloy and temperature;
alloying and ordering can change lattice heat transport
nonmonotonically~\cite{wang2021alloythermal}.

\section{Two-dimensional transport normalization}
\label{sec:si-normalization}
For supercell height $c$ and effective material thickness $t$, the
normalization definitions in the main article give
\begin{align}
\sigma_\alpha&=\frac{c}{t}\,\sigma_{\mathrm{SC},\alpha},\\
\kappa_{\mathrm e,\alpha}
&=\frac{c}{t}\,\kappa_{\mathrm e,\mathrm{SC},\alpha},\\
n_{\mathrm{2D}}&=c\,n_{\mathrm{SC}}=t\,n.
\end{align}
Here, SC denotes quantities normalized to the complete periodic
supercell volume. The height $c$ includes the material and vacuum;
it is not the vacuum width alone. For $n_{\mathrm{SC}}$ in
$\mathrm{cm^{-3}}$, $c$ must be expressed in cm to obtain
$n_{\mathrm{2D}}$ in $\mathrm{cm^{-2}}$. The Seebeck coefficient
requires no thickness rescaling.

The corresponding sheet conductances are
$G_{\square,\alpha}=c\sigma_{\mathrm{SC},\alpha}=t\sigma_\alpha$ and
$K_{\square,\mathrm e,\alpha}
=c\kappa_{\mathrm e,\mathrm{SC},\alpha}=t\kappa_{\mathrm e,\alpha}$.
The lattice reference must use the same thickness convention,
$K_{\square,\mathrm l}^{\mathrm{ref}}=t\kappa_{\mathrm l}^{\mathrm{ref}}$,
when combined with the electronic terms in $ZT$~\cite{wu2017sheet}.
Thickness invariance requires the physical sheet conductances to remain
fixed. Changing $t$ while keeping the numerical lattice conductivity
fixed at $550~\mathrm{W\,m^{-1}\,K^{-1}}$ instead changes the reference
model. The effective thickness is defined from the relaxed-bilayer
interlayer separation as stated above; the complete supercell height
$c$ remains a separate quantity.

\section{Quantum-confinement schematic}
\begin{figure}[htbp]
    \centering
    \includegraphics[width=0.99\linewidth]{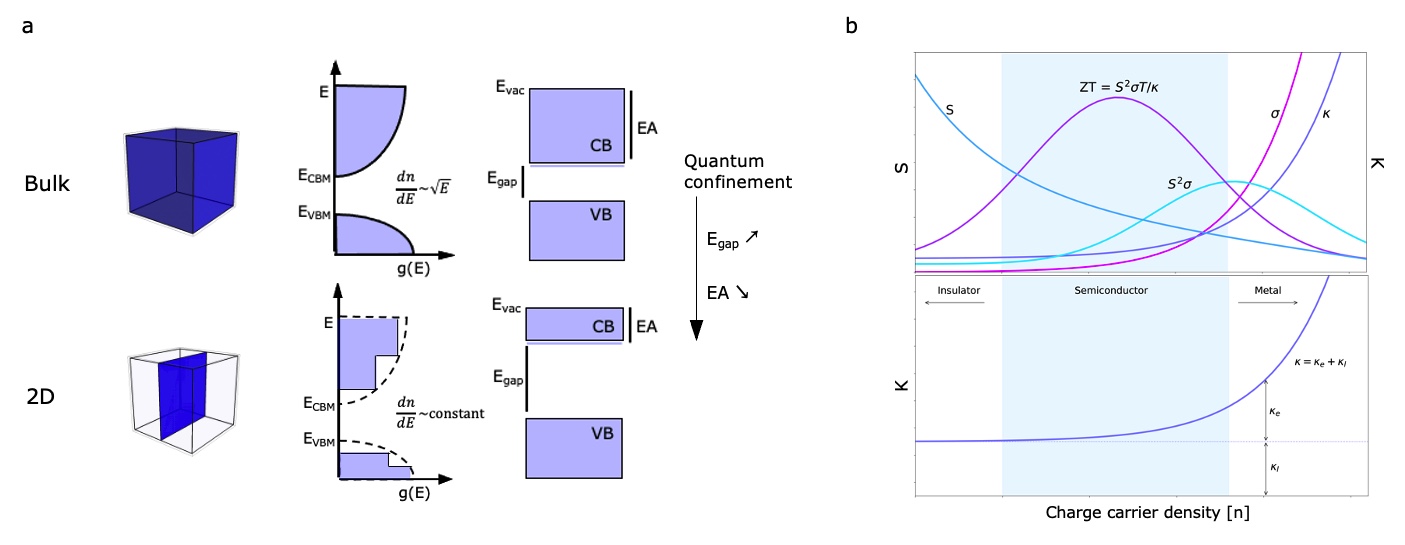}
    \caption{Qualitative schematics of (a) the effect of quantum confinement
    on the density of states and band structure of semiconductor materials
    and (b) the dependence of the thermoelectric figure of merit $ZT$,
    Seebeck coefficient $S$, electrical conductivity $\sigma$, and thermal
    conductivity $\kappa$ on carrier concentration.}
    \label{fig:QC}
\end{figure}

Figure~\ref{fig:QC} provides qualitative context for the dimensionality and
carrier-concentration effects discussed in the main article.

\clearpage
\section{Estimated intrinsic dielectric breakdown field}
\label{sec:si-breakdown}

The intrinsic dielectric breakdown fields of the monolayers were estimated
using the phenomenological model of Kim, Pilania, and
Ramprasad~\cite{kim2016breakdown}, which combines the electronic band gap
$E_{\mathrm g}$ and the maximum phonon frequency $\nu_{\max}$:
\begin{equation}
\frac{E_{\mathrm b}}{\mathrm{MV\,cm^{-1}}}
=0.24442\exp\!\left[
0.315\sqrt{
\frac{E_{\mathrm g}}{\mathrm{eV}}
\frac{\nu_{\max}}{\mathrm{THz}}}
\right].
\label{eq:si-breakdown}
\end{equation}
This is Eq.~(3) of Ref.~\cite{kim2016breakdown}, converted from
$\mathrm{MV\,m^{-1}}$ to $\mathrm{MV\,cm^{-1}}$; the prefactor in the
original unit convention is $24.442$. Here, $\nu_{\max}$ is the ordinary
phonon cutoff frequency, corresponding to the quantity denoted
$\omega_{\max}$ in that reference. Its numerical value is entered in THz
without a factor of $2\pi$. It is distinct from the dipole-weighted effective
polar-optical-phonon frequency $\omega_{\mathrm{eff}}$ defined in
Sec.~\ref{sec:si-computational-methods}. The band gap and maximum frequency
describe electronic excitation and phonon-mediated energy relaxation,
respectively. The supplied maximum-frequency data are interpreted as
wavenumbers $\widetilde{\nu}_{\max}$ in $\mathrm{cm^{-1}}$ and converted using
$\nu_{\max}/\mathrm{THz}=0.0299792458\,
(\widetilde{\nu}_{\max}/\mathrm{cm^{-1}})$.
Table~\ref{tab:si-breakdown-data} lists the wavenumbers, converted
frequencies, and breakdown fields to two decimal places, matching the
field precision in main-article Table~I(a).

\begin{figure}[!htbp]
    \centering
    \includegraphics[width=0.80\linewidth]{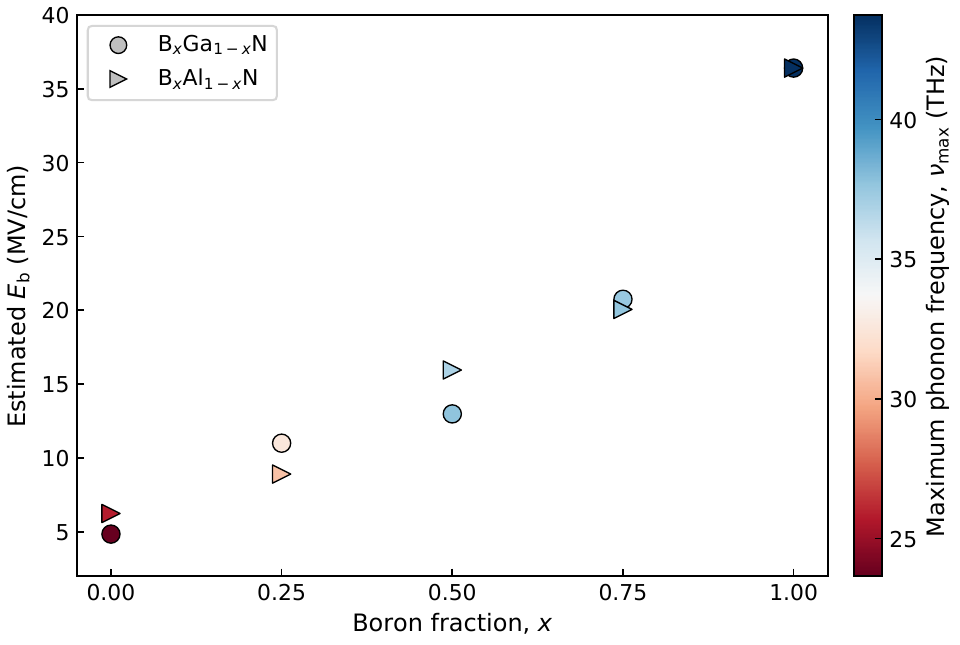}
    \caption{Estimated intrinsic dielectric breakdown field $E_{\mathrm b}$
    of monolayer \chem{B_xGa_{1-x}N} (circles) and
    \chem{B_xAl_{1-x}N} (triangles) as a function of boron fraction $x$,
    from the numerical outputs of Eq.~\ref{eq:si-breakdown} listed in
    Table~\ref{tab:si-breakdown-data}. The color scale denotes the
    maximum phonon frequency $\nu_{\max}$ in THz after wavenumber
    conversion. The two series share the hBN endmember at $x=1$;
    its two markers overlap.}
    \label{fig:si-breakdown}
\end{figure}

Figure~\ref{fig:si-breakdown} shows an increase in the estimated breakdown
field with boron content, from $6.23$ and $4.84~\mathrm{MV\,cm^{-1}}$
for hAlN and hGaN, respectively, to
$36.40~\mathrm{MV\,cm^{-1}}$ for the common hBN endpoint. The maximum
phonon frequency likewise rises overall toward the B-rich end.
Within Eq.~\ref{eq:si-breakdown}, larger values of either
$E_{\mathrm g}$ or $\nu_{\max}$ increase $E_{\mathrm b}$. The relative
ordering of the two series changes at intermediate compositions.
At $x=0.5$ the Ga-containing alloy has the higher cutoff
frequency ($37.76$ versus $36.91$~THz) but the lower estimated field
($12.98$ versus $15.95~\mathrm{MV\,cm^{-1}}$).
Thus, the cutoff frequency alone is insufficient to rank the estimated
breakdown fields.

The B-rich alloys therefore exhibit high predicted intrinsic breakdown
fields. For context, Slobodyan \textit{et al.}~\cite{slobodyan2022critical}
assessed bulk-diode critical fields at a common drift-region doping of
$10^{16}~\mathrm{cm^{-3}}$, obtaining approximately $3.4~\mathrm{MV/cm}$
for GaN and $10~\mathrm{MV/cm}$ for diamond; the widely quoted
$8~\mathrm{MV/cm}$ for $\beta$-\chem{Ga_2O_3} was identified as an
estimate. These comparisons provide a scale for the present predictions;
differences in dimensionality, field profiles, and methods preclude a
direct ranking of device performance.

\newpage
The Kim model was fitted to first-principles intrinsic-breakdown data for
bulk crystalline insulators~\cite{kim2016breakdown}. Its transfer to these
monolayers requires validation, and its scalar form does not resolve field
direction. The hBN endpoint illustrates this limitation: the present model
gives $36.40~\mathrm{MV/cm}$, whereas Hattori \textit{et al.} measured
approximately $3~\mathrm{MV/cm}$ in-plane ($E\perp c$) and
$12~\mathrm{MV/cm}$ out-of-plane ($E\parallel c$) in single-crystal
hBN under controlled humidity~\cite{hattori2016anisotropic}.
Those bulk-crystal measurements and the monolayer estimate describe
different geometries. Direction-resolved calculations or measurements are
needed to establish attainable fields; defects, contacts, and the device
environment also affect experimental breakdown. The present results thus
support further investigation of high-field applications.
These estimates complement the main article's thermoelectric results but
do not enter $ZT$ or replace the state-dependent ADP-plus-PIE transport model.

\begin{table}[!htbp]
\caption{Maximum phonon wavenumbers, corresponding frequencies, and
estimated intrinsic dielectric breakdown fields, rounded to two decimal
places.}
\label{tab:si-breakdown-data}
\centering
\small
\begin{tabular*}{\textwidth}{@{\extracolsep{\fill}}lcccc}
\toprule
Material & $x$ &
\shortstack{$\widetilde{\nu}_{\max}$\\($\mathrm{cm^{-1}}$)} &
\shortstack{$\nu_{\max}$\\(THz)} &
\shortstack{Estimated $E_{\mathrm b}$\\(MV/cm)} \\
\midrule
hAlN & 0 & 862.01 & 25.84 & 6.23 \\
\chem{B_{0.25}Al_{0.75}N} & 0.25 & 1026.86 & 30.78 & 8.90 \\
\chem{B_{0.5}Al_{0.5}N} & 0.5 & 1231.13 & 36.91 & 15.95 \\
\chem{B_{0.75}Al_{0.25}N} & 0.75 & 1255.40 & 37.64 & 20.06 \\
\addlinespace[3pt]
hGaN & 0 & 788.99 & 23.65 & 4.84 \\
\chem{B_{0.25}Ga_{0.75}N} & 0.25 & 1086.10 & 32.56 & 10.99 \\
\chem{B_{0.5}Ga_{0.5}N} & 0.5 & 1259.42 & 37.76 & 12.98 \\
\chem{B_{0.75}Ga_{0.25}N} & 0.75 & 1250.36 & 37.48 & 20.74 \\
\addlinespace[3pt]
hBN & 1 & 1459.14 & 43.74 & 36.40 \\
\bottomrule
\end{tabular*}
\end{table}

\clearpage
\linespread{1.35}\selectfont
\bibliographystyle{apsrev4-2}
\bibliography{references}